      [1999/12/01 v1.4c Il Nuovo Cimento]
\documentclass{cimento}
\usepackage{graphicx}
\title{GRB-supernovae: a new spin on gravitational waves}
\author{Maurice H.P.M.~van Putten\from{ins:x}}
\instlist{\inst{ins:x} 
    MIT-LIGO Project, NW 17-161, 175 Albany Street, Cambridge, MA 02139}
\PACSes{\PACSit{98.70.Rz}{98.62}}
\begin{document}

\maketitle

\begin{abstract}
The discovery of the GRB-supernova association 
poses the question on the nature of the inner
engine as the outcome of Type Ib/c supernovae. These events are believed
to represent core-collapse of massive stars, probably in low-period stellar
binaries and similar but not identical to the Type II event SN1987A. 
The branching ratio of Type Ib/c supernovae into GRB-supernovae has the 
remarkably small value of less than 0.5\%. These observational constraints 
point towards a rapidly rotating black hole formed at low probability with
low kick velocity. The putative black hole hereby remains centered, and
matures into a high-mass object with large rotational energy in angular momentum.
As the MeV-neutrino emissions from SN1987A demonstrate, the most powerful probe 
of the inner workings of core-collapse events are radiation channels to which 
the remnant envelope is optically thin. We here discuss the prospect of gravitational-wave
emissions powered by a rapidly rotating central black hole which, in contrast to
MeV-neutrinos, can be probed to distances as large as 100Mpc through upcoming
gravitational-wave detectors LIGO and Virgo. We identify the GRB-emissions, commonly attributed
to ultrarelativistic baryon-poor ejecta, with a new process of linear 
acceleration of charged particles along the axis of rotation of a black hole in 
response to spin-orbit coupling. 
We include some preliminary numerical simulations on internal 
shocks produced by intermittent ejecta. The results showing a radial splash, which 
points towards low-luminosity and lower-energy radiation at large angles,
possibly related to X-ray flashes.
\end{abstract}

\section{Introduction}

The Type Ib/c GRB-supernova association, beautifully demonstrated by
GRB030329/ SN2002dh, establishes SNIb/c as the parent population of
cosmological gamma-ray bursts. This, in turn, confirms their 
association to massive stars proposed by Woosley \cite{woo93,kat94,pac98}.
Massive stars have characteristically short lifetimes, whereby GRB-supernovae
track the cosmological event rate. This is quantitatively 
confirmed by agreement in their
true-to-observed event rates of 450-500 based on two independent analysis,
one on geometrical beaming factors and the other on locking to the 
cosmic star-formation rate \cite{fra01,van03}.

Furthermore, Type II and Type Ib/c supernovae are believed to represent the endpoint of 
massive stars \cite{fil97,tur03} in binaries, such as in the Type II/Ib event 
SN1993J \cite{mau04}. This binary association suggests a hierarchy, wherein
hydrogen-rich, envelope retaining SNII are associated with wide binaries, 
while hydrogen-poor, envelope stripped SNIb and SNIc are associated with 
increasingly compact binaries \cite{nom95,tur03}. By tidal coupling,
the primary star in the latter will rotate at approximately the obital period
at the moment of core-collapse. With an evolved core \cite{bro00}, these events
are therefore believed to produce a spinning black hole \cite{woo93,lee02,bet03}. 

GRB-supernovae are rare, in view of the observed small branching ratio of less 
than 1\% off their parent SNIb/c. While all SNIb/c
may be producing black holes, only some are associated with the
production of ultrarelativistic outflows powering gamma-ray emissions
\cite{ree92,ree94}. Quite generally, black holes produced in core-collapse
possess random kick velocities of a few hundred km~s$^{-1}$ \cite{bek73}.
Quite generally, therefore, black holes formed in core-collapse are parametrized
by their mass $M$, angular momentum $J$ and kick velocity $K$. By its kick
velocity, a newly formed black hole typically escapes the central high-density 
regions of the progenitor star before core-collapse is completed. This suggests 
an association of GRB-supernovae with the small sample of black holes 
possessing low kick velocities \cite{van04b}. In these cases, the black hole 
remains centered, surges into a high-mass object with a rotational energy 
of up to 66\% of the maximal attainable value, and subsequently spins up by 
accretion or spins down in a suspended accretion state \cite{van04b}.

Spacetime around a rotating black hole is described by the Kerr metric 
\cite{ker63}. It describes an exact solution of frame-dragging induced 
by the angular momentum of a black hole. Frame-dragging appears in 
nonzero angular velocities of zero angular momentum observers. Observational
evidence for frame-dragging rotating black holes is successfully pursued 
through X-ray spectroscopy by ASCA and XMM in surrounding accretion disks 
\cite{tan95,iwa96,fab04}.
Kerr black holes further possess an anomalously large energy reservoir in
angular momentum. Per unit mass, this energy reservoir exceeds that of
a rapidly spinning neutron star by at least an order of magnitude. Finally,
this energy reservoir is entirely baryon-free. 

Frame-dragging creates a large differential angular velocity between the
black hole, when rapidly spinning, and surrounding matter. When the latter
is unmagnetized, the black hole continues to accrete matter, enlarging it 
and spinning it up towards an extreme Kerr black hole along a Bardeen 
trajectory \cite{bar72}. Alternatively, when the surrounding matter is 
uniformly magnetized, corresponding to two counter-oriented current rings 
\cite{van03b}, a spin-connection becomes effective between the
angular momentum of the black hole and the inner regions of the surrounding
accretion disk which forcefully spins down the black hole \cite{van04b}.
Especially at high spin rates, this process is facilitated by an equilibrium
magnetic moment of the black hole in its lowest energy state.
This process is self-sustaining, as most of the black-hole luminosity is
incident onto the inner face of a surrounding torus, created with strong
differential shear by competing torques acting on its inner and outer face.
This energy input may be the agent driving a dynamo which creates
superstrong magnetic fields. In a suspended accretion state,
the large energy reservoir in black hole spin energy is released with
an efficiency of $\eta=\Omega_T/\Omega_H$ determined by the ratio of 
the angular velocity $\Omega_T$ of the surrounding torus to the angular 
velocity $\Omega_H$ of the black hole. This energy release, on the order of 
$10^{53}$erg. This output is larger than that seen at present in the 
electromagnetic radiation of GRB-emissions and supernova ejecta 
(about $10^{51}$ erg, corrected for beaming and asphericities).  
However, this additional energy output can be safely accounted for by
``unseen" emissions in gravitational radiation and MeV-neutrino emissions 
\cite{van01,van03b,van04a}.

Frame-dragging also acts along the axis of rotation. Here, it creates an 
energetic interaction along open magnetic flux-tubes (``ergo-tubes") subtended 
by the event horizon of the black hole. These interactions define a new 
mechanism of accelerating magnetized ejecta to ultrarelativistic
velocities. This spin-orbit coupling is a classical analogue of an
earlier quantum mechanical treatment \cite{van00}. 
This application radically differs from the common view, that 
black-hole energetic processes are strictly confined to the action of 
frame-dragging in the ergosphere. 
We shall illustrate the case of intermittent ejecta with a numerical 
simulation, whose shocked interactions serve as input to baryon-poor outflows powering 
gamma-ray emissions \cite{ree92,ree94}.

\section{Theoretical background}

An extreme Kerr black hole reaches an angular
velocity of its horizon given by $\Omega_H=1/2M$. This angular velocity far
surpasses that of surrounding matter, when separated from the event horizon
by a distance on the order of the linear size of the black hole. According to
the first law of thermodynamics, the efficiency of energy transfer from the 
black hole to the environment reaches a fraction 
\begin{eqnarray}
\eta=\frac{\Omega_T}{\Omega_H}
\label{EQN_ETA}
\end{eqnarray}
of its rotational energy $E_{rot}=2M\sin^2(\lambda/4)$, where $\sin\lambda$ 
the specific angular momentum of the black hole per unit mass and $\Omega_T$
denotes the angular velocity of the surrounding torus. The torus itself is
suspended by competing torques acting on its two faces, one positive torque on its
inner face and a negative torque on its outer face due to generic losses in energy
and angular momentum in magnetic winds. The question arises: what are the
radiation channels through which the torus processes this input? We shall discuss
this in the next section.

The Kerr metric shows black hole spin creating frame-dragging, described by
particular contributions to the Riemann tensor \cite{cha83}. In turn,
the Riemann tensor couples to spin \cite{pap51a,pap51b,pir56,mis74}.
Rotating black holes hereby couple to the angular momentum (and spin) of 
nearby particles. We note that the angular momentum per unit mass 
represents a rate of change of surface area, while the Riemann tensor has 
dimension cm$^{-2}$ in geometrical units. Hence, curvature-spin coupling produces 
a force, whereby test particles follow non-geodesic trajectories \cite{pir56}.
These gravitational spin-angular momentum interactions are commonly referred to 
as gravitomagnetic effects \cite{tho86} by analogy to magnetic moment-magnetic 
moment interactions, even though there is an interesting difference in sign. To 
quantify this in the application to outflows from black holes, we shall focus on 
spin-orbit interactions along the axis of rotation following \cite{van00}. 
Using dimensional analysis once more, the gravitational potential for spin aligned 
interactions should satisfy
\begin{eqnarray}
E=\omega J,
\label{EQN_USS}
\end{eqnarray}
where $\omega$ refers to the frame-dragging angular velocity produced
by the massive body and $J$ is the angular momentum of the particle.
Spinning bodies hereby couple to spinning bodies \cite{oco72}.

We shall detail the radiative process of (\ref{EQN_ETA}) and (\ref{EQN_USS}) 
in case of a particular topology of the magnetosphere, consisting of a torus
magnetosphere supported by two counter-oriented current rings and, formed by
a changes of topology from the outer layers of the inner torus magnetosphere, 
an open magnetic flux tube along the axis of rotation \cite{van01,van03b}.

\section{Radiation channels of a non-axisymmetric torus}

The surrounding matter is strongly sheared by competing torques, which creates
a torus with a super-Keplerian inner face and a sub-Keplerian outer face. 
For sufficiently strong viscosity -- regulated by magnetohydrodynamical stresses -- between
the inner and the outer face, a state of suspended accretion can arise, in which
the energy and angular momentum input from the black hole is balanced by energy
and angular momentum losses to infinity. 
This {\em suspended accretion state} is described by balance equations of 
energy and momentum flux, and in many ways is equivalent
to pulsars when viewed in poloidal topology \cite{van03b}.
The kinetic energy of the torus intruduces a bound on the poloidal magnetic
field-energy that it can support. We recently derived an estimate for the
a maximal ratio of poloidal magnetic field 
energy-to-kinetic energy of the torus of about 1/12 \cite{van03b}. 

Rapidly rotating black holes dissipate most of their spin-energy
in the event horizon, set essentially by the spin-rate of the black hole.
The lifetime of rapid spin hereby satisfies \cite{van03b}
\begin{eqnarray}
T_s\simeq 90\mbox{s} \left(\frac{M_H}{7M_\odot}\right)\left(\frac{\eta}{0.1}\right)^{-8/3}
\left(\frac{\mu}{0.03}\right)^{-1}.
\label{EQN_TS}
\end{eqnarray}
This estimate is consistent with durations of tens of seconds of long gamma-ray 
bursts \cite{kou93}.

The gravitational wave-emissions due to quadrupole emissions
by a mass-inhomogeneity $\delta M_T$ are \cite{pet63}
\begin{eqnarray}
L_{gw}=\frac{32}{5}\left(\omega{\cal M}\right)^{10/3}F(e)
       \simeq\frac{32}{5}\left(M_H/R\right)^5(\delta M_T/M_H)^2,
\label{EQN_PET}
\end{eqnarray}
where $\omega\simeq M^{1/2}/R^{3/2}$ denotes the orbital frequency of the torus with
major radius $R$, ${\cal M}=(\delta M_T M_H)^{3/5}/(\delta M_T+M_H)^{1/5}\simeq
M_H(\delta M_T/M_H)^{5/3}$ denotes the chirp mass\index{Chirp mass}, 
and $F(e)$ denotes a geometric
factor representing the ellipticity $e$ of the orbital motion.
The linearized result (\ref{EQN_PET}) has been amply confirmed by approximately
one-solar luminosity in gravitational waves emitted by
PSR1913+16\index{PSR1913+16} (with ellipticity $e=0.62$ \cite{hul75,tay94}). 
Here, we apply (\ref{EQN_PET}) to a non-axisymmetric torus
around a black hole,
whose mass-quadrupole inhomogeneity $\delta M_T$ is determined self-consistently in
a state of suspended accretion for the lifetime of rapid spin of the black hole. 
Quadrupole mass-moments will appear spontaneously due to non-axisymmetric waves
whenever the torus is sufficiently slender. 

The competing torques on the inner and outer face
promote azimuthal shear in the torus, leading to a super-Keplerian
state of the inner face and a sub-Keplerian state of the outer face. 
The torus becomes geometrically thick and may hereby become sufficiently slender
to allow instability of quadrupole wave-modes $m=2$ for
a minor-to-major radius less than 0.3260 \cite{van03b}. This 
produces gravitational radiation at close to 
twice the angular frequency of the torus.

The equations of suspended accretion \cite{van03b} can be solved
algebraically, giving solutions for the total
radiation energies emitted by the torus, expressed as fractions of
the rotational energy of the black hole. Solving for the simultaneous
output in in gravitational radiation, MeV-neutrino emissions and 
magnetic winds, we find
\begin{eqnarray}
E_{gw}\simeq 0.2M_\odot\left(\frac{\eta}{0.1}\right)
\left(\frac{M_H}{7M_\odot}\right),
~~~f_{gw}\simeq 500\mbox{Hz}
   \left(\frac{\eta}{0.1}\right)
   \left(\frac{7M_\odot}{M_H}\right)
\label{EQN_1GW}
\end{eqnarray}
in units of $M_\odot=2\times 10^{54}$erg. This appreciation
(\ref{EQN_1GW}) of GRBs typically surpasses the true energy $E_\gamma\simeq 3\times 10^{50}$ 
erg in gamma-rays \cite{fra01} by several orders of magnitude with
potentially some exceptions \cite{ghi04}.
Sub-dominant emissions in magnetic winds power an accompanying supernova
at a factor $\eta$ less than that in gravitational radiation,
\begin{eqnarray} 
E_{w}=4\times10^{52}\mbox{erg}\left(\frac{\eta}{0.1}\right)^2
\left(\frac{M}{7M_\odot}\right).
\label{EQN_EW}
\end{eqnarray}

The energy in torus winds (\ref{EQN_EW}) further provide a powerful agent towards various processes:
an accompanying supernovae \cite{bet03}, possibly radiatively-driven and
radio loud by dissipation of their magnetic energy \cite{van04a},
collimation of the enclosed baryon-poor outflows from the black hole \cite{lev00}, 
as well as a source of neutron pick-up by the same \cite{lev03}.

The energy output in thermal and MeV-neutrino emissions
is a factor $\delta$ less than that in gravitational radiation, or
\begin{eqnarray} 
E_{\nu}=2\times 10^{53}\mbox{erg}\left(\frac{\eta}{0.1}\right)
                               \left(\frac{\delta}{0.30}\right)
                               \left(\frac{M_H}{7M_\odot}\right).
\label{EQN_EN}
\end{eqnarray}
At this dissipation rate, the torus develops a temperature of a few MeV 
and produces baryon-rich winds \cite{van03b}.

\section{Spin-orbit interactions $E=\Omega J$}

A particle in a periodic orbit, e.g., a charge particle
in a Landau state confined to a surface of constant magnetic flux, 
carries orbital angular momentum. Geometrically, this motion represents
a rate of change of surface area an associated spacelike two-surface, 
following the dimensional analysis mentioned in the previous section. 
This two-surface can be identified by considering the helical motion as
seen in four-dimensional spacetime \cite{van05}. The result is an
interaction with the Riemann tensor -- the Papapetrou interaction between
spin or angular momentum with curvature \cite{pap51a,pap51b}.

The Kerr metric embodies an exact solution to spin-induced curvature,
expressed by the appearance of frame-dragging in the
Riemann-curvature. Not surprisingly, the Papapetrou
spin-curvature coupling produces spin-coupling to the frame-dragging
angular velocity. 
In a neighborhood of the axis of rotational of the black hole, this gives
rise to (\ref{EQN_USS}). To see this, we may use
the non-zero components of the Riemann tensor of the Kerr metric 
in Boyer-Lindquist coordinates relative to tetrad 1-forms
given by Chandrasekhar \cite{cha83}. 
Accordingly, the Riemann tensor creates the radial force \cite{van05}
\begin{eqnarray}
F_2=JR_{3120}=JAD=-\partial_2\omega J.
\label{EQN_QQ6}
\end{eqnarray}
The assertion (\ref{EQN_USS}) follows from
\begin{eqnarray}
E=\int_r^\infty F_2 ds = \omega J.
\label{EQN_USS1}
\end{eqnarray}
The result (\ref{EQN_USS1}) also follows from a completely independent derivation, 
by considering the
difference in total energy between two particles in counter rotating orbits about 
the axis of rotation of the black hole. 
We insist that these two particles have angular momenta of opposite sign and 
equal magnitude, $J_\pm = g_{\phi\phi} u^t (\Omega_\pm - \omega)$, where
$u^b$ denotes their velocity four-vectors. Thus, we have
\begin{eqnarray}
J_\pm =g_{\phi\phi}u^t 
 \sqrt{\omega^2 - (g_{tt}+(u^t)^{-2})/g_{\phi\phi}}=\pm J.
\label{EQN_EJ}
\end{eqnarray}
This shows that $u^t$ is the same for each particle. The total
energy of the particles is given by $E_\pm = (u^t)^{-1}+\Omega_\pm J_\pm,$
and hence one-half their difference \cite{van05}
\begin{eqnarray}
E=\frac{1}{2}(E_+-E_-)=\omega J.
\label{EQN_USS2}
\end{eqnarray}

A perfectly conducting blob of charged particles in a magnetic field in 
electrostatic equilibrium is characterized by rigid rotation \cite{tho86} 
with angular velocity $\Omega_b$. 
In the frame of zero angular momentum observers, 
the local charge-density is given by the Goldreich-Julian charge density 
\cite{gol69}. In the lowest energy state with vanishing canonical 
angular momentum, the angular momentum of the charge particles satisfies 
$J=eA_\phi$, where $e$ denotes the unit of electric charge and $A_\phi$ the 
$\phi-$component of the electromagnetic vector potential $A_a$ \cite{van00}. 

Consider blob of charged particles in an open magnetic
flux tube about the axis of rotation of the black hole with 
magnetic flux $2\pi A_\phi$. 
The number $N(s)$ of particles per unit height $s$ of the blob, therefore, 
satisfies $N(s)=(\Omega_b-\omega)A_\phi/e,$ where $e$ denotes the elementary charge.
A pair of blobs in both directions along the spin-axis of scale height $h$ 
hereby receives an energy
\begin{eqnarray}
E_{blob}= \omega JNh = \omega(\Omega_b-\omega)A^2_\phi h=
\left(1\times 10^{47}\mbox{erg}\right)B_{15} h_M^3 H,
\label{EQN_BL}
\end{eqnarray}
where $h_M=h/M$ denotes the linear dimension of the blob, $B_{15}=B/10^{15}$.
Here, we use the normalized function
$H=4\hat{\omega}\left(\hat{\Omega}_b-\hat{\omega}\right)$,
where $\hat{\omega}=\omega/\Omega_H$ and $\hat{\Omega}_b=\Omega_b/\Omega_H$.
Charged particles in superstrong magnetic fields are essentially massless.
The ejection of a pair of blobs with energy (\ref{EQN_BL}) hereby takes place in
a light-crossing time of about $0.3$ms for a stellar mass black hole, such as a 
seven solar mass black hole of linear dimension $10^7$cm. This corresponds to an 
instantaneous luminosity on the order of $3\times 10^{50}$erg/s.

\begin{figure}
\centerline{\includegraphics[scale=0.50]{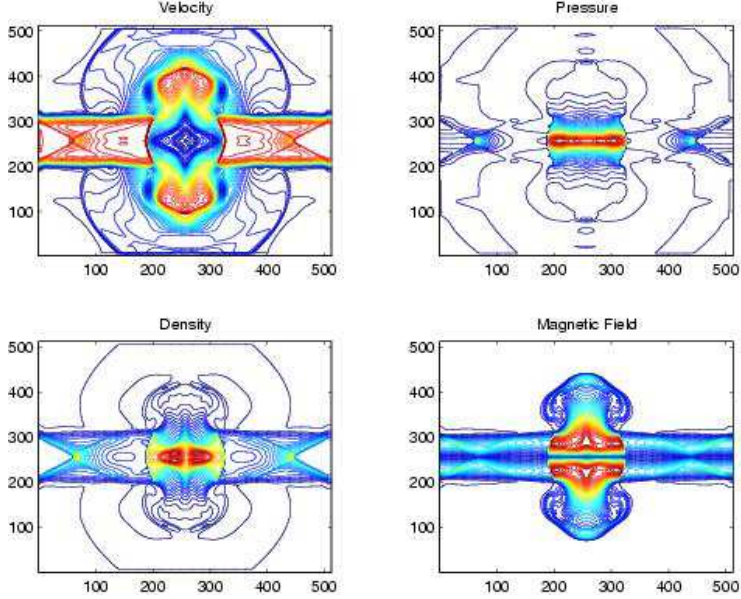}}
\caption{The head-on two magnetized ejecta (``kissing jets") in the comoving frame 
of the center of mass, representing the collision of a fast blob overtaking a slow 
blob at large distances from an intermittent source. The simulation by JETLAB F90 
has incoming flows each with Lorentz factor 1.5, and
densities a few times that of the unmagnetized environment.
A high-energy density region forms about a comoving
stagnation point with strong amplification of the transverse magnetic field,
and a radial splash produced by a pressure-driven, subsonic and radially outgoing 
flow with accompanying cocoon. This transverse morphology points towards accompanying 
low-luminosity low-energy gamma-ray or X-ray emissions over large angles.}
\label{FIG_54}
\end{figure}

An intermittent source produces iterated emissions of blobs -- ``pancakes" in
the observer's frame \cite{pir04} -- which can dissipate their kinetic energy 
in collisions in the internal shock model of gamma-ray bursts \cite{ree92,ree94}.
Fig. (\ref{FIG_54}) shows a relativistic simulation of such internal shock using
JETLAB F90.

This presentation in terms of blobs is complementary to discussions on continuous 
outflows \cite{lev04}, wherein gamma-ray emissions are attributed 
to shocks due to by steepening by temporal fluctuations or intermittency at the 
source \cite{lev97}, or late-time interaction with the environment.

\section{GRB-supernovae as long-duration burst sources for LIGO/Virgo}

\begin{figure}
\begin{minipage}[t]{6.5cm}
\includegraphics[width=0.9\textwidth]{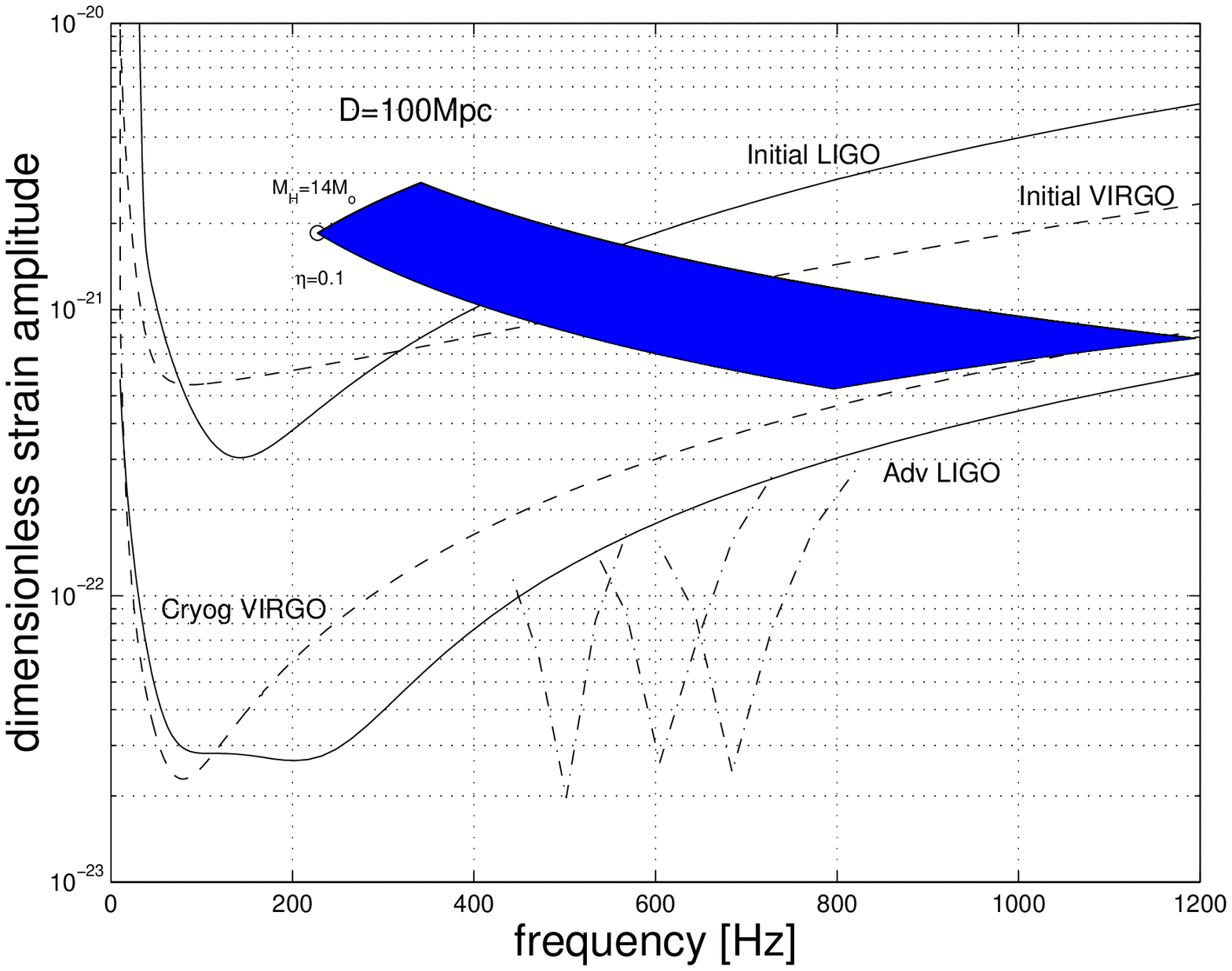} 
\caption[]
{The expected distribution of strain amplitudes of GRB-supernovae 
and the design targets for the strain
noise amplitude of the LIGO and Virgo detectors in the
$h(f)$-diagram. The fiducial source distance is 100Mpc,
corresponding event rate is about one per year.}
\label{FIG:fp1}
\end{minipage}
\hfill
\begin{minipage}[t]{6.5cm}
\includegraphics[width=0.9\textwidth]{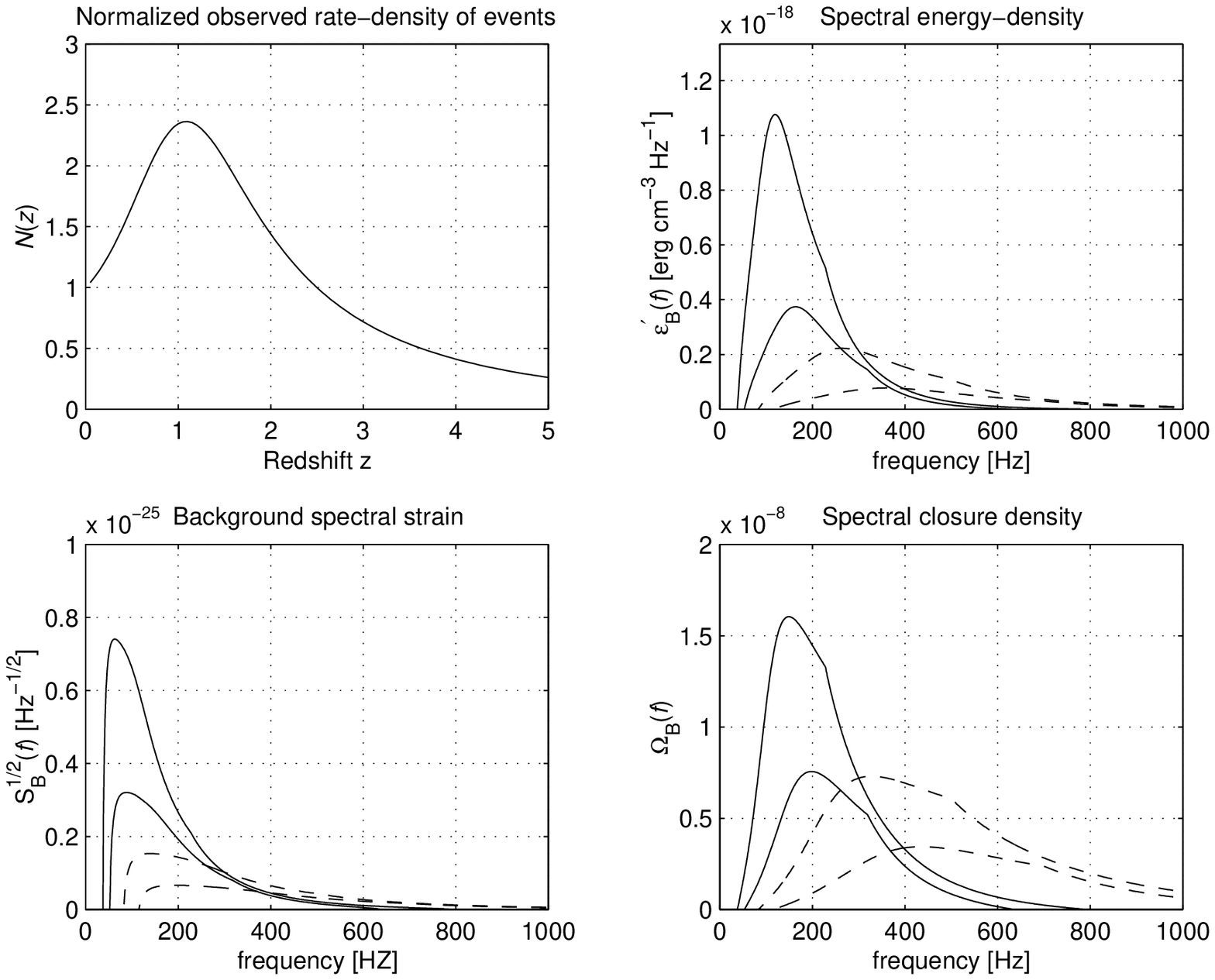} 
\caption[]{The expected cosmological distribution of GRB-supernovae
to the stochastic background in gravitational radiation, on the basis
of their correlation to the cosmic star-formation rate $N(z)$.
(Reprinted from \cite{van05}\copyright Cambridge University Press.)}
\label{FIG:fp2}
\end{minipage}
\hfill
\end{figure}

The predicted gravitational radiation (\ref{EQN_1GW}) is in the
range of sensitivity of the broad band detectors LIGO \cite{abr92,bar99} 
and Virgo \cite{bra92,ace02,ale04} shown in Fig. (\ref{FIG:fp1}), as 
well as GEO \cite{dan95,wil02} and TAMA \cite{and02}.
Matched filtering gives a theoretical upper bound on the signal-to-noise 
ratio in the detection of the long bursts in gravitational radiation from
GRB-supernovae, shown in Fig. (\ref{FIG:fp1}. 
As a hydrodynamical source, the frequency will be unsteady
at least on an intermittent timescales, so that a time-frequency trajectory 
method is probably more applicable in practice \cite{van04a}. 
The signal-to-noise ratio will hereby be between first and second-order
detection algorithms. 

As the outcome of Type Ib/c supernovae, GRB-SNe are an astrophysical source 
population locked to the cosmic star-formation rate. Thus, their
contribution to the stochastic background in gravitational radiation can be
estimated semi-analytically, given the band-limited signals assuming
$B=\Delta f/f_e$ of around 10\%, where $f_e$ denotes the average
gravitational-wave frequency in the comoving frame. Here, we can use
the scaling relations
$E_{gw} = E_0 M_H/M_0,~~~f_e = f_0 M_0/M_H$
where $M_0=7M_\odot$, $E_0=0.203M_\odot \eta_{0.1}$ and
$f_0=455$Hz$\eta_{0.1}$, assuming maximal spin-rates 
($E_{rot}=E_{rot,max}$). For non-extremal black holes, a
commensurate reduction factor in energy output can be inserted.

Assuming a uniform distribution of black hole masses,
e.g., $M_H=(4-14)\times M_\odot$, uncorrelated to
the relative angular velocities $\eta$ of the torus, 
summation gives an expected spectral energy-density \cite{van04a}
\begin{eqnarray}
<\epsilon_B^\prime(f)> = 1.08 \times 10^{-18} \hat{f}_B(x)
\mbox{~erg~cm}^{-3}~\mbox{Hz}^{-1}
\end{eqnarray}
where $\hat{f}_B(x) = f_B(x)/$max$f_B(\cdot)$ is a normalized
frequency distribution. The associated dimensionless amplitude
$\sqrt{S_B(f)}=\sqrt{2G/\pi c^3} f^{-1} \tilde{F}_B^{1/2}(f)$, where
$\tilde{F}_B=c\epsilon_B^\prime$ and $G$ denotes Newton's constant.
satisfies
\begin{eqnarray}
\sqrt{S_B(f)}=7.41\times 10^{-26}\eta_{0.1}^{-1} \hat{f}_S^{1/2}(x)
\mbox{~Hz}^{-1/2}
\end{eqnarray}
where $\hat{f}_S(x) = f_S(x)/$max$f_S(\cdot)$, $f_S(x)=f_B(x)/x^2$,
Likewise, we have for the spectal closure density $\Omega_B(f)
=f \tilde{F}_B(f)/\rho_cc^3$ relative to the closure density
$\rho_c = 3H_0^2/8\pi G$
\begin{eqnarray}
\tilde{\Omega}_B(f)= 1.60\times 10^{-8} \eta_{0.1} \hat{f}_\Omega(x),
\end{eqnarray}
where $\hat{f}_\Omega(x) = f_\Omega(x)/$max$f_\Omega(\cdot)$,
$f_\Omega(x)=xf_B(x)$ and $H_0$ denotes the Hubble constant. 

The results are shown in Figs. (\ref{FIG:fp1}-\ref{FIG:fp2}).
The extremal value of $\Omega_B(f)$ is in the
neighborhood of the location of maximal sensitivity of LIGO and
Virgo.  

{\bf Acknowledgment.}
The author thanks A. Levinson, E. Schuryak, R. Preece, A. Levinson,
R.P. Kerr for their constructive comments. This research is
supported by the LIGO Observatories, constructed by Caltech and MIT
with funding from NSF under cooperative agreement PHY 9210038.
The LIGO Laboratory operates under cooperative agreement
PHY-0107417. 


\begin{thebibliography}{0}
\bibitem{abr92}\BY{Abramovici, A., Althouse, W.E., Drever, R.W.P., et al.}
\IN{Science}{292}{1992}{325}
\bibitem{ace02}\BY {Acernese, F., et al.}\IN{Class. Quant. Grav.}{19}{2002}{1421}
\bibitem{and02}\BY {Ando, M., and the TAMA Collaboration}{Class. Quant. Grav.}{19}{200}{1409}
\bibitem{bar72}\BY {Bardeen, J.M., Press, W.H., \& Teukolsky, S.A.}
 \IN{ApJ}{178}{1972}{347}  
\bibitem{bar99}\BY {Barish, B., \& Weiss, R.} \IN{Phys. Today}{52}{1999}{44}
\bibitem{bra92}\BY{Bradaschia, C., Del Fabbro, R., di Virgilio, A., {et al.}}, 
 \IN{Phys. Lett. A}{163}{1992}{15}
\bibitem{bek73}\BY{Bekenstein, J.D.}\IN{ApJ}{183}{1973}{657}
\bibitem{bet03}\BY{Bethe, H.A., Brown, G.E., \& Lee, C.-H.}
 \TITLE{Selected Papers: Formation 
 and Evolution of Black Holes in the Galaxy}{ (World Scientific, 2003)} 
\bibitem{bro00}\BY{Brown, G.E., Lee, C.-H., Wijers R.A.M.J., Lee, H.K., Israelian G. 
 \& Bethe H.A.}\IN{NewA}{5}{2000}{191}
\bibitem{cha83}\BY{Chandrasekhar, S.}
 \TITLE{The Mathematical Theory of Black Holes},
 (Oxford: Oxford University Press, 1983)
\bibitem{dan95}\BY{Danzmann, K.}
 \TITLE{in First Edoardo Amaldi Conf. Grav. Wave Experiments}, 
 {E. Coccia, G. Pizella, F. Ronga (Eds.)}{ (Singapore: World Scientific, 1995)}, p100
\bibitem{fab04}\BY{Fabian, A.C.}
  \TITLE{From X-ray Binaries to Quasars: Black Hole Accretion on All Mass Scales}, 
  {ed. T.J. Maccarone, R.P. Fender, and L.C. Ho}{ (Dordrecht: Kluwer, 2004)}{}
\bibitem{fil97}\BY{Filipenko, A.V.}\IN{ARA\&A}{35}{1997}{309}
\bibitem{fra01}\BY{Frail, D.A., Kulkarni, S.R., Sari, R., et al.} 
 \IN{ApJ}{562}{2001}{L55}
\bibitem{ghi04}\BY{Ghirlanda et al.}\IN{ApJ}{616}{2004}{331}
\bibitem{gol69}\BY{Goldreich, P., \& Julian, W.H.}\IN{ApJ}{157}{1969}{869} 
\bibitem{hul75}\BY{Hulse, R.A., \& Taylor, J.H.}\IN{ApJ}{1975}{195}{L51} 
\bibitem{iwa96}\BY{Iwasawa, K., Fabian, A.C., Reynolds, C.S., et al.}\IN{MNRAS}
	                {282}{1996}{1038}
\bibitem{kat94}\BY{Katz, J.I.}\IN{ApJ}{432}{1994}{L27}
\bibitem{ker63}\BY{Kerr, R.P.}\IN{Phys. Rev. Lett.}{11}{1963}{237}
\bibitem{kou93}\BY{Kouveliotou, C., Meegan, C.A., Fishman, G.J., {et al.}}
 \IN{ApJ}{413}{1993}{L101}
\bibitem{lan84}\BY{Landau, L.D., \& Lifschitz, E.M.}
 \TITLE{Classical Theory of Fields}{ (Oxford: Pergamon Press, 1984)}
\bibitem{oco72}\BY{O'Connel, R.F.}\IN{Phys. Rev. D.}{10}{1974}{3035} 
\bibitem{lee02}\BY{Lee, C.-H., Brown, G.E., \& Wijers, R.A.M.J.}\IN{ApJ}{575}{2002}{996}
\bibitem{lev97}\BY{Levinson A., van Putten, M.H.P.M.}\IN{ApJ}{488}{1977}{69}
\bibitem{lev00}\BY{Levinson, A., \& Eichler, D.}\IN{Phys. Rev. Lett.}{2000}{85}{236}
\bibitem{lev03}\BY{Levinson, A., \& Eichler, D.}{ApJ}{594}{2003}{L19} 
\bibitem{lev04}\BY{Levinson, A.}\IN{ApJ}{608}{2004}{411}
\bibitem{mau04}\BY{Maund, J.R., Smartt, S.J., Kudritzki, R.P., Podsiadiowski, P.,\& Gilmore, G.F.}
 \IN{Nature}{427}{2004}{129}
\bibitem{mis74}\BY{Misner, Thorne, K.S., Wheeler, A.}
 \TITLE{Gravitation (San Francisco)} 1974
\bibitem{nom95}\BY{Nomoto, K., Iwamoto, K., \& Suzuki, T}\IN{Phys. Rep.}{256}{1995}{173}
\bibitem{pap51a}\BY{Papapetrou, A.}\IN{Proc. Roy. Soc.}{209}{1951}{248}
\bibitem{pap51b}\BY{Papapetrou, A.}\IN{Proc. Roy. Soc.}{209}{1951}{259}
\bibitem{pac98}\BY{Paczy\'nski, B.P.}\IN{ApJ}{494}{1998}{L45}
\bibitem{pet63}\BY{Peters, P.C., and Mathews, J.}
 \IN{Phys. Rev.}{131}{1963}{435}
\bibitem{pir04}\BY{Piran, T.}\IN{astro-ph/0405503}{}{2004}
\bibitem{pir56}\BY{Pirani, F.A.E.}\IN{Act. Phys. Pol.}{XV}{1956}{389} 
\bibitem{ree92}\BY{Rees, M.J., \& M\'esz\'aros, P.}\IN{MNRAS}{258}{1992}{41P}
\bibitem{ree94}\BY{Rees, M.J., \& M\'esz\'aros, P.}\IN{ApJ}{430}{1994}{L93}
\bibitem{ale04}\BY {Spallici, A.D.A.M., Aoudia, S., de Freitas Pacheco, J.A., 
 et al.}\IN{gr-qc archive}{2004}{gr-qc/0406076}
\bibitem{tan95}\BY{Tanaka, Y., Nandra, K., Fabian, A.C.}\IN{Nature}{375}{1995}{659}
\bibitem{tay94}\BY{Taylor, J.H.}\IN{Rev. Mod. Phys.}{66}{1994}{711}
\bibitem{tho86}\BY{Thorne, K.S., Price, R.H., MacDonald, D.A.}
 \TITLE{Black Holes:The Membrane Paradigm}
 { (New Haven: Yale University Press, 1986)}
\bibitem{tur03}\BY{Turatto, M.}\TITLE{Supernovae and Gamma-ray Bursters} 
 {ed. K.W. Weiler}{ (Heidelberg: Springer-Verlag, 2003)}, p21
\bibitem{van99}\BY{van Putten, M.H.P.M.}\IN{Science}{284}{1999}{115}
\bibitem{van00}\BY{van Putten, M.H.P.M.}\IN{Phys. Rev. Lett.}{84}{2000}{3752}
\bibitem{van03}\BY{van Putten, M.H.P.M., \& Regimbau, T.} 
 \IN{ApJ}{593}{2003}{L15}
\bibitem{van04a}\BY{van Putten, M.H.P.M., Levinson, A., Lee, H.-K., Regimbau, T., \& Harry, G.}
 \IN{Phys. Rev. D.}{69}{2004}{044007}
\bibitem{van04b}\BY{van Putten, M.H.P.M.}\IN{ApJ}{611}{2004}{L81} 
\bibitem{van03b}\BY{van Putten, M.H.P.M., \& Levinson, A.}\IN{ApJ}{584}{2003}{953} 
 \bibitem{van01}\BY{van Putten, M.H.P.M.}\IN{Phys. Rev. Lett.}{87}{2001}{091101}
\bibitem{van05}\BY{van Putten, M.H.P.M.}
\TITLE{Gravitational Radiation, Luminous Black Holes and Gamma-ray Burst
Supernovae}{ (Cambridge: Cambridge University Press, 2005)}
\bibitem{wil02}\BY{Willke, B., et al.}{Class. Quant. Grav.}{19}{2002}{1377}
\bibitem{woo93}\BY{Woosley, S.E.}\IN{ApJ}{405}{1993}{273}
\end{thebibliography}
\end{document}